\documentclass[aps,pra,reprint,superscriptaddress]{revtex4-1}
\usepackage[T1]{fontenc}

\usepackage[english]{babel} 
\usepackage{wasysym}
\usepackage{graphicx}
\usepackage{shadow}
\usepackage{epsfig}
\usepackage{natbib}
\usepackage{dcolumn}
\usepackage{hyperref}
\usepackage{float}
\hypersetup{
backref=true,
pagebackref=true,
hidelinks
}
\newcommand{\vect}[1]{\vec{#1}}

\begin{document}
\title{Universality of spin-relaxation for spin 1/2 particles diffusing over magnetic field inhomogeneities in the adiabatic regime}
\author         {M. Guigue}
\email          {guigue@lpsc.in2p3.fr}
\affiliation{LPSC, Universit\'e Grenoble-Alpes, CNRS/IN2P3, Grenoble, France}
\author {R. Golub}
\email {rgolub@ncsu.edu}
\affiliation {Physics Department, North Carolina University, Raleigh, NC 27965}
\author         {G. Pignol}
\email          {pignol@lpsc.in2p3.fr}
\affiliation{LPSC, Universit\'e Grenoble-Alpes, CNRS/IN2P3, Grenoble, France}
\author{A. K. Petukhov}
\email{petukhov@ill.fr}
\affiliation{Institut Laue Langevin, 6, Rue Jules Horowitz, 38000 Grenoble}
\date{\today}

\begin{abstract}
We present a theoretical analysis of spin relaxation, for a polarized gas of spin 1/2 particles undergoing restricted adiabatic diffusive motion within a container of arbitrary shape, due to magnetic field inhomogeneities of arbitrary form. 
\end{abstract}

\maketitle

\section{Introduction}

Hyperpolarized gas such as $^{3}$He is used in a wide variety of scientific and medical situations. 
In physics, $^{3}$He is commonly used as a spin-filter for neutrons \cite{Andersen2006}, as a precision magnetometer \cite{Gemmel2010} and as a probe for new fundamental spin-dependent  interactions \cite{Petukhov2010,Fu2011,Petukhov2011,Bulatowicz2013,Tullney2013}. 
In medicine, it is used for magnetic resonance imaging \cite{VanBeek2003a,Thien2008,Zheng2011b}.
In certain conditions, the polarization lifetime of a $^{3}$He cell can be as long as a few weeks, which is a desirable feature in most of those applications. 
To reach and improve on low depolarization rates, all possible depolarization phenomena should be carefully controlled. 
One important depolarization channel is the relaxation induced by the motion of the polarized particles in an inhomogeneous magnetic field. 

This depolarization has been of ongoing interest since the pioneering work in the field \cite{Bloembergen1948,Bouchiat1960,Gamblin1965,Schearer1965}. 
The polarization evolution is characterized by three quantities: the longitudinal and transversal depolarization rate $\Gamma_{1}$ and $\Gamma_{2}$ and the associated frequency shift $\delta\omega$. 
In many experimental setups, the Helium 3 polarized gas is contained in cells with typical sizes of $10\,\mathrm{{cm}}$ at pressure of the order of $1\,\mathrm{{bar}}$. 
In those cases, the gas is in the diffusive regime, where interparticle collisions are more frequent than wall collisions. 

While there are various theoretical approaches \cite{Grebenkov2007},  we will treat the depolarization rate induced by magnetic field inhomogeneities  using the results of the standard perturbation theory \cite{Redfield1965,McGregor1990,Goldman2001,Lamoreaux2005b,Petukhov2010}.
This has been applied in various ways in the literature; the Redfield approach \cite{Redfield1965, Slichter1963, McGregor1990} starting with the equation of motion of the density matrix has been shown to be equivalent to the perturbation theory applied to the Torrey equation for the density matrix by expansion in eigenfunctions \cite{Cates1988,Golub2010c}, as well as equivalent to a perturbative solution of the same equation based on the Green's function and a direct perturbative solution of the Schroedinger equation \cite{Golub2014}.
Since the approach is  perturbative,  it is valid only for a limited range of appropriate parameters.

\begin{figure}[ptbh]
\centering
\includegraphics[scale=0.045,angle=90]{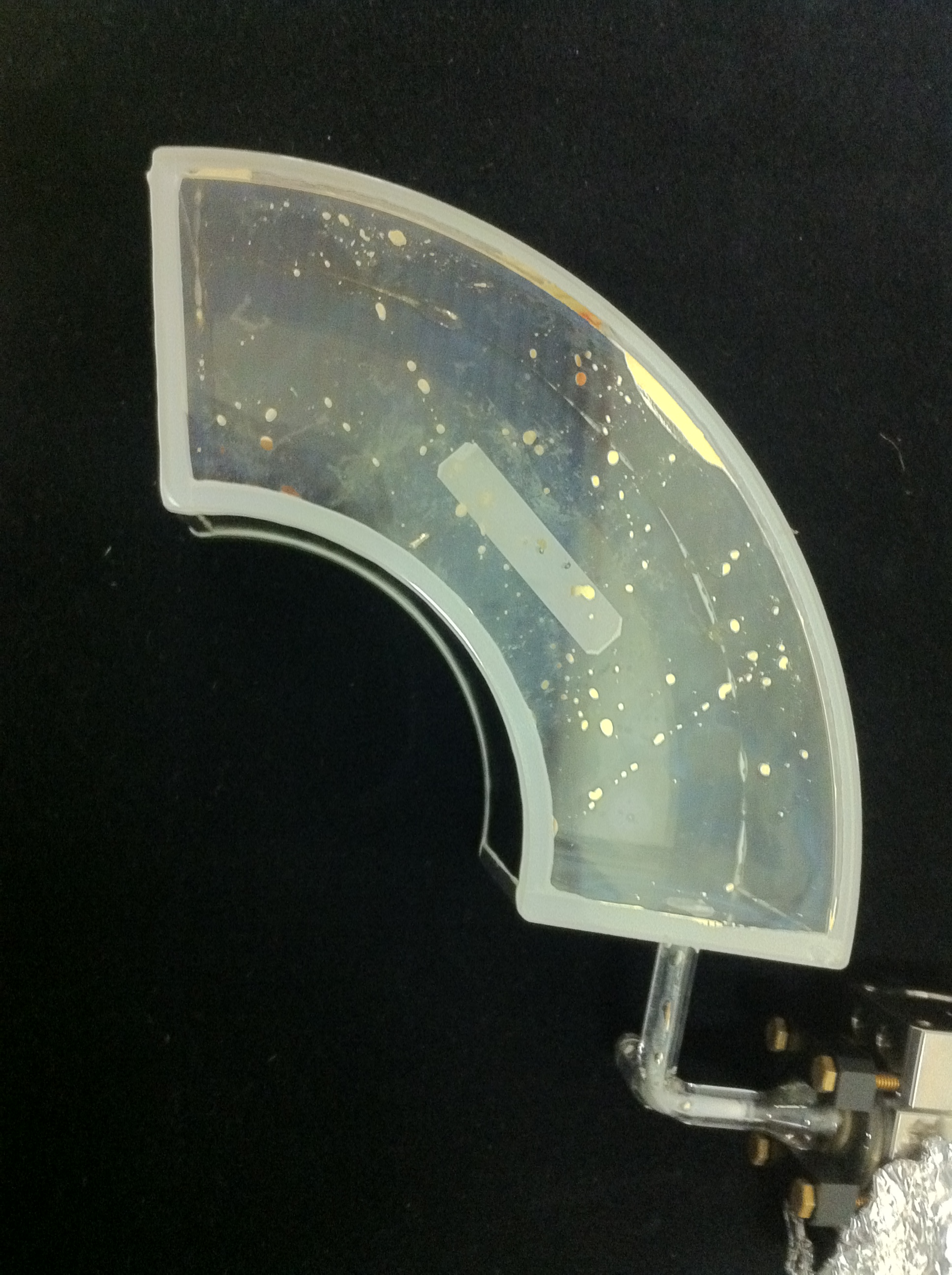}
\includegraphics[scale=0.045]{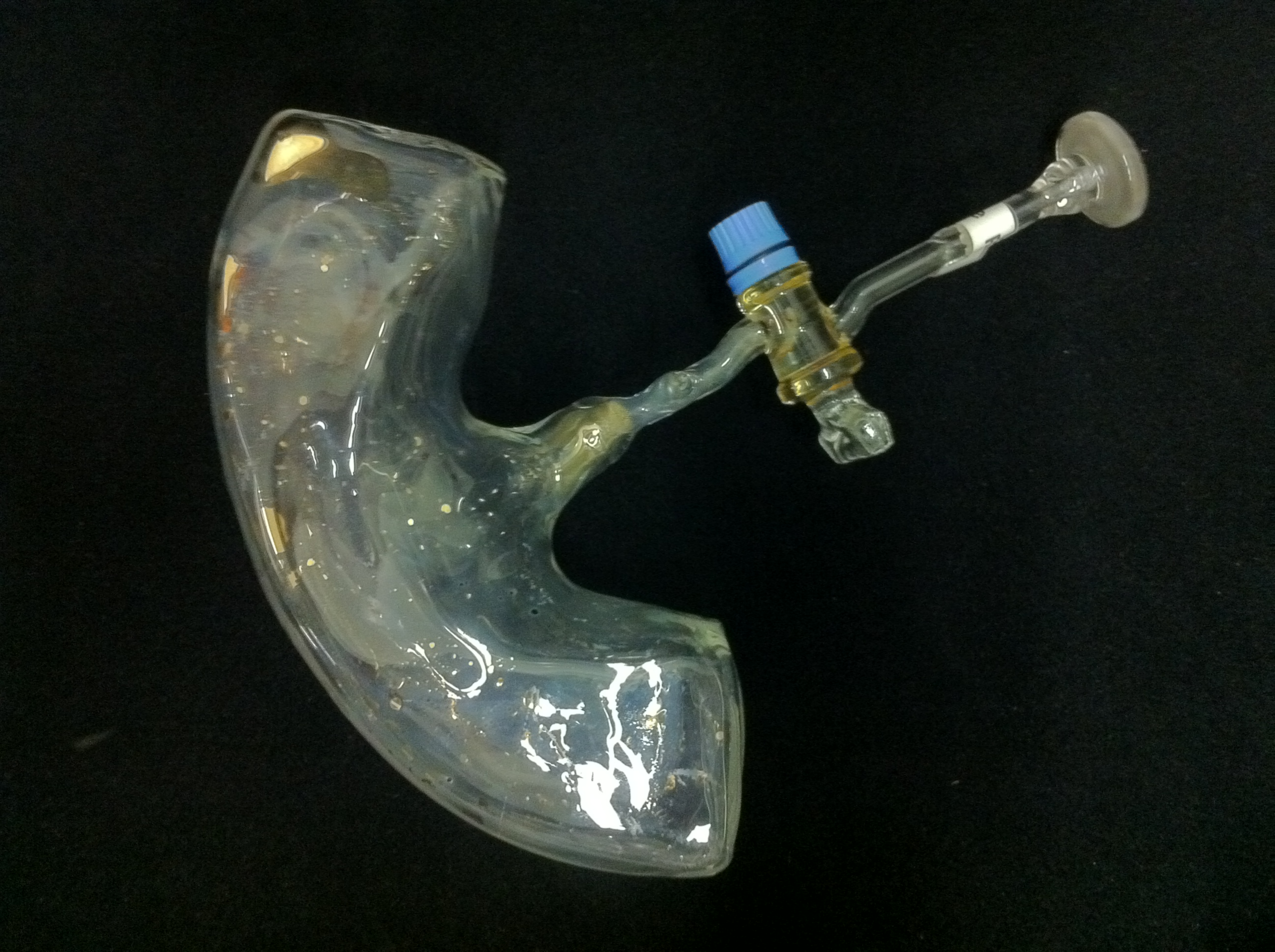}
\includegraphics[scale=0.065]{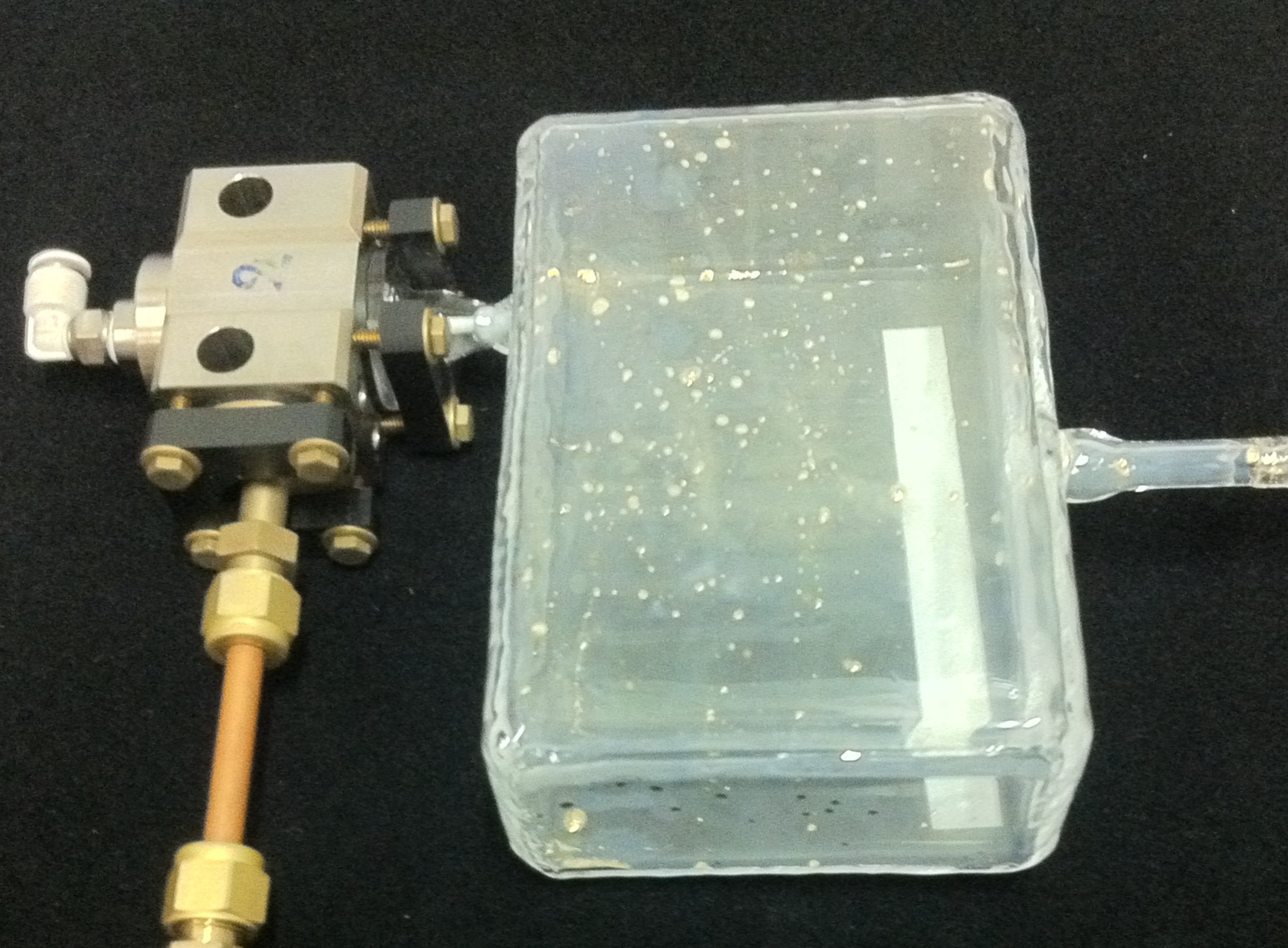}
\includegraphics[scale=0.045]{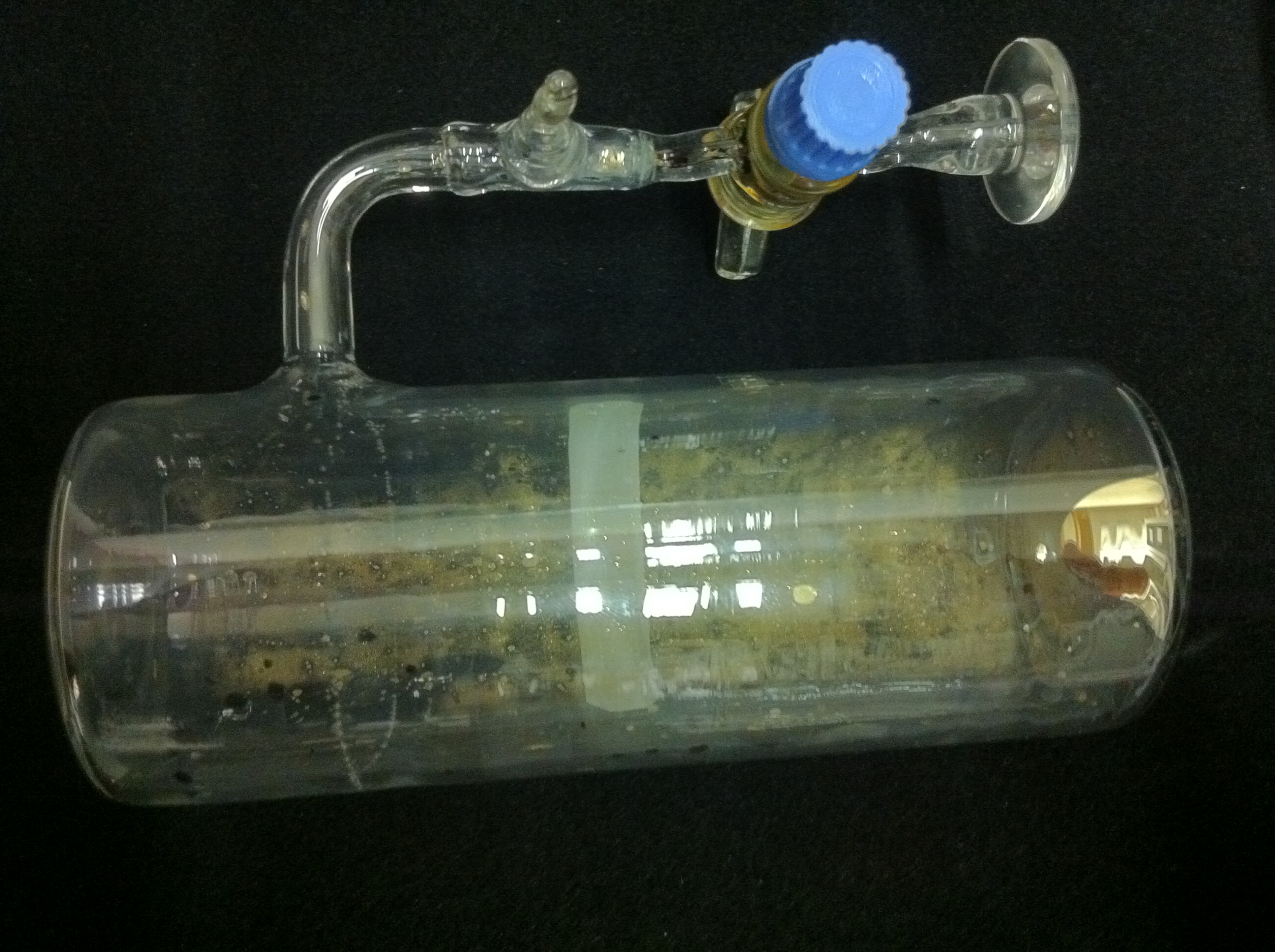}
\caption{Examples of cells used as polarized helium 3 containers.}
\label{fig:cells}
\end{figure}

The early theoretical treatments of the problem were limited to simple cell geometries (1D, cylindrical or spherical in \cite{McGregor1990} ) and the magnetic field  {gradients}  {were} assumed uniform over the cell volume. 
However, every-day practice often has to deal with magnetic field  inhomogeneities with a second or even higher order variation with position. 
The situation is even more complex if one uses  cells with very special shapes (such as wide-angle "banana" cells \cite{Stewart2006}; see Fig. \ref{fig:cells} for examples of cells used nowadays for polarized helium 3).

Very recently, a more general solution valid for an arbitrary magnetic field has been proposed in terms of a Fourier series expansion \cite{Petukhov2010} for a 1D diffusive motion. 
A 3D generalization for the case of an arbitrary magnetic field and a rectangular cell  {valid for all values of interparticle collision rate is given } in \cite{Clayton2011,Swank2012}.

In this paper we present simple analytical expressions for $\Gamma_{1}$ and $\delta\omega$ valid for spin 1/2 particles undergoing adiabatic restricted diffusive motion (the term "adiabatic" means that the holding magnetic field is high, $\omega _{0}\tau_{\rm{corr}}\gg 1$,  {where }$\omega_{0}=\gamma B_{0}$  {is the Larmor frequency and }$\tau_{\rm{corr}}$   is the correlation time for the field fluctuations seen by the particles, and the spins approximately follow its local direction: this will be discussed in detail in Section IV) within a cell of arbitrary form and influenced by a magnetic field of arbitrary shape, using the Redfield theory involving correlation functions. 
The conditions necessary to apply the Redfield  (perturbation) theory and our result will be clearly explained. 
Also an exact solution for power law ($b[z]\approx gz^{k},k=1,2,3,4$) inhomogeneity is presented and compared with our approximate result.

\section{General concepts of polarized gas relaxation}

We will consider the case of an assembly of spin one-half particles with a gyromagnetic ratio $\gamma$ evolving in a slightly inhomogeneous magnetic field.

For a polarized gas at  "normal conditions" contained in a typical 10 cm cell exposed to a weak (few tenths of Gauss) magnetic field with a tiny inhomogeneity across the cell, the field correlation time $\tau _{\rm{corr}}$ is approximately the time constant of the lowest diffusion mode $\tau _{\rm{corr}} \approx \tau _D \approx R^2/D\pi ^2 \approx 1\,\rm{s}$.
The total magnetic field in which the gas is immersed is defined as $\vec{B}=B_{0}\vec{e}_{z}+\vec{b}$ where $\vec{b}=b_{x}\vec{e}_{x}+b_{y}\vec{e}_{y}+b_{z}\vec{e}_{z}$ only depends on the position. 
The magnetic field inhomogeneity corresponds to $\vec{b}$. 
We defined $\omega_{0}=\gamma B_{0}$ and $\langle b_{i}\rangle=0$.

To apply the Redfield theory for spin-relaxation in slightly inhomogeneous magnetic fields, the resulting relaxation time $T$ must be longer than the correlation time $\tau_{\rm{corr}}$ \cite{Redfield1965,Goldman2001}. 
If one is looking for the longitudinal relaxation rate $\Gamma _1$, then this time $T$ is $T_1 =1/\Gamma _{1}$; for the frequency shift, it  is equal to $1/\delta\omega$.
This condition implies that the relaxation is slow enough to let the spins diffuse across the cell many times before they are strongly relaxed. 
Typical values for magnetic inhomogeneities gradients are $\frac{\vec{\nabla}b}{B_{0} }\approx10^{-3}$ to $10^{-4}\,\mathrm{{cm}^{-1}}$, leading to longitudinal spin-relaxation time constants from a few tenths to a few thousands hours, so in practice, the condition of applicability of Redfield  (perturbation) theory is well fulfilled.

The attractive feature of Redfield theory is the result that expresses the observables of interest $\Gamma _1$, $\Gamma _2$ and $\delta  \omega$ in terms of the Fourier spectra of corresponding field correlation functions.
\begin{widetext}
\begin{equation}\label{eq:long-relaxation-rate}
\Gamma _1=\frac{1}{T_1}=\gamma ^2  \left(\mathcal{R}e  \left[S_{xx}(\omega _0) + S_{yy}(\omega _0)\right] +\mathcal{I}m\left[ S_{yx}(\omega _0) -S_{xy}(\omega _0)\right]\right),
\end{equation}
\begin{equation}\label{eq:trans-relaxation-rate}
\Gamma _2=\frac{1}{T_2} =\frac{1}{2} \Gamma _1 +\gamma ^2  S_{zz}(0),
\end{equation}
\begin{equation}\label{eq:frequency-shift}
\delta \omega =\frac{\gamma ^2}{2}\left(\mathcal{R}e \left[ S_{xy}(\omega _0) - S_{yx}(\omega _0)\right] + \mathcal{I}m \left[ S_{xx}(\omega _0) +S_{yy} (\omega _0)\right]\right),
\end{equation}
\end{widetext}
where $S_{ij}(\omega)$ is the Fourier transform or spectrum of the magnetic field correlation function defined as:
\begin{equation}
S_{ij}(\omega)=\int_{0}^{\infty}\langle b_{i}(0)b_{j}(\tau)\rangle\exp
(i\omega\tau)\mathrm{{d}\tau}.%
\end{equation}
The ensemble average of the variable $X$ is denoted  {by} $\langle X\rangle$. 
The correlation function of $b_{i}$ and $b_{j}$ can be expressed as:
\begin{equation}
\langle b_{i}(0)b_{j}(\tau)\rangle=\frac{1}{V}\int_{V}\mathrm{{d}}\vec{r}_{0}\int_{V}\mathrm{{d}}\vec{r}b_{i}(\vec{r}_{0})b_{j}(\vec{r})p(\vec{r},\tau\mid\vec{r}_{0})\label{defn:corr-functions},%
\end{equation}
where $V$ is the volume of the cell and $p(\vec{r},\tau\mid\vec{r}_{0})$ is the conditional probability (or propagator) for a particle which is at $\vec{r}_{0}$ at $t=0$ to be at $\vec{r}$ at the time $t$.

Moreover, $p(\vec{r},\tau\mid\vec{r}_{0})$  {satisfies} the initial condition:
\begin{equation}
p(\vec{r},t=0\mid\vec{r}_{0})=\delta(\vec{r}-\vec{r}_{0} )\label{eq:initial-condition}.
\end{equation}
As a consequence, we have:
\begin{equation}
\langle b_{i}(0)b_{j}(0)\rangle=\frac{1}{V}\int_{V}\mathrm{{d}}\vect{r}b_{i}(\vec{r})b_{j}(\vec{r})=\overline{b_{i}b_{j}}.
\label{def:volume-average}
\end{equation}
$\overline{X}$ corresponds to the volume average of the quantity $X$.

When the gas is in the diffusion regime, (the mean free path between interparticles collisions is  {much }shorter than the mean free path between wall collisions);  the propagator is governed by the diffusion equation:
\begin{equation}
\frac{\partial p(\vec{r},\tau\mid\vec{r}_{0})}{\partial\tau}=D\bigtriangleup
p(\vec{r},\tau\mid\vec{r}_{0})\label{eq:diffusion},
\end{equation}
where $D$ is the diffusion coefficient which is inversely proportional to the pressure of the gas. 
This equation gives correct solutions  for the time behavior of the propagator,  {for times}  longer than $\tau _{ \rm{coll}}$, the mean time between particle collisions.

We only consider the relaxation  due to restricted diffusive spin motion in a slightly inhomogeneous magnetic field. 
Any additional depolarization due to possible interactions of the spin with magnetic impurities in/on the cell walls is neglected. 
This leads to the boundary condition on the container walls \cite{Cates1988}:
\begin{equation}
\vec{\nabla}p(\vec{r},\tau|\vec{r}_{0})\cdot \vec{n}=0\label{eq:boundaries},%
\end{equation}
where $\vec{n}$ is the vector normal to the surface.

The shape of the magnetic field inhomogeneity $b$ must be taken into account in order to calculate the correlation functions (\ref{defn:corr-functions}). 
For the simplest case of an uniform gradient and specific cell geometry (rectangular, spherical, cylindrical \cite{Cates1988,McGregor1990,Clayton2011}), it is commonly known that at high magnetic field and pressure, the longitudinal relaxation rate can be expressed as:%
\begin{equation}
\label{eq:commonlyknowresult}
\Gamma_{1}=D\frac{|\vec{\nabla}b_{x}|^{2}+|\vec{\nabla}b_{y}|^{2}}{B_{0}^{2}}.%
\end{equation}
For higher orders of power law inhomogeneities and when the gas is confined in a rectangular cell, the solution has been obtained in the form of a Fourier series \cite{Petukhov2010,Swank2012} (For details, see Appendix C).
Similar behavior is observed for all power law inhomogeneities at high magnetic field and pressure (see Fig. 2). 
But for more complicated cell geometries, the solution of the problem is quickly limited by one's capability of finding an appropriate basis, in which the propagator can be expanded. \begin{figure}[h]
\centering
\includegraphics[scale=0.7]{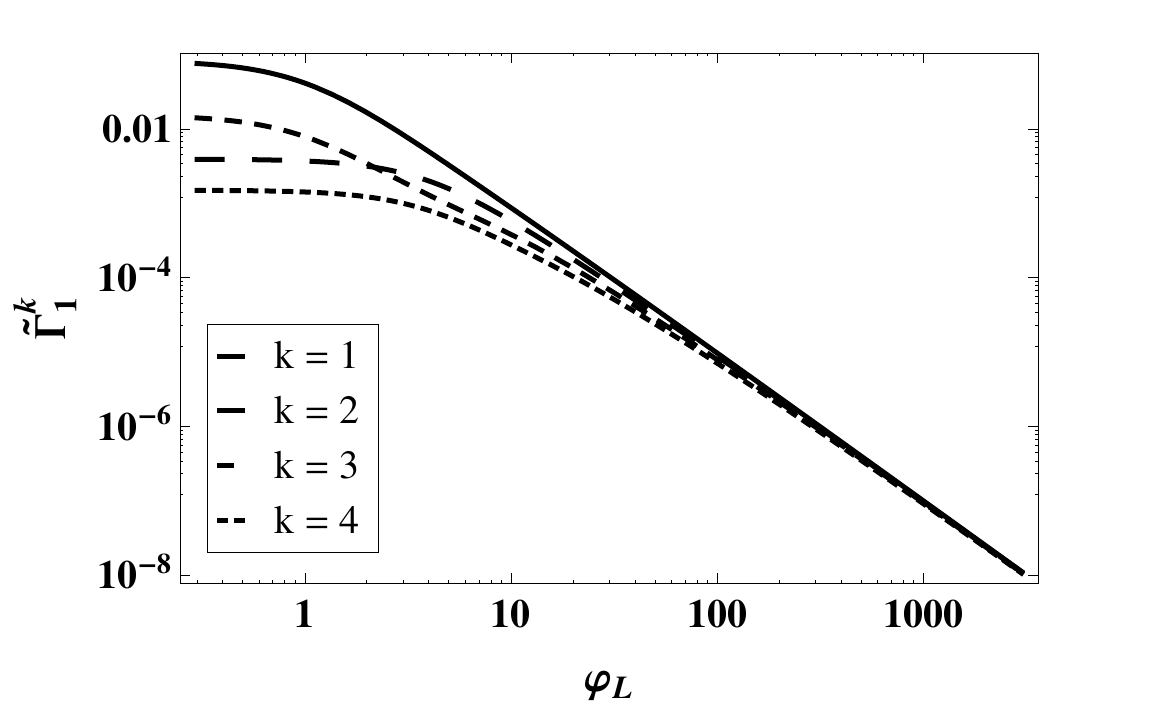}
\caption{Behavior of normalized relaxation rates $\tilde{\Gamma}_{1}^{k}$ defined with Eq. (\ref{eq:gamma_tild}) for different power law magnetic field inhomogeneities depending on $\phi_{L}=\gamma B_{0}\frac{L^{2}}{\pi^{2}D}$.}
\label{fig:comb_plot}%
\end{figure}

\section{Longitudinal relaxation rate in the diffusive adiabatic regime}

From Eq. (\ref{eq:long-relaxation-rate}), we see that the longitudinal relaxation rate $\Gamma_{1}$ is defined by the spectrum of the magnetic field correlation functions. 
Below, we show that for high frequencies the spectrum, and hence relaxation rate, can be obtained in a closed form for arbitrary magnetic inhomogeneities  {and} container shapes.

Further on, we will be interested in a time scale much longer than a typical time between particle collisions $\tau _{\rm{coll}}$.
For such times $\left\{ t_k \right\}$, the field $\vec{b}(t_k)$ experienced by the spin of a randomly moving particles forms a stationary stochastic process $\left\{ b_j (k)\right\}$.
This assumption leads us to the time translational invariance of the field correlation function:
\begin{equation}\label{eq:timeinv_corrfunction}
\langle b_j (t) b_j (t+\tau)\rangle = \langle b_j (0) b_j (\tau)\rangle .
\end{equation}
In addition, for the field $\left\{b _i\right\}$ dependent explicitly only on the position, the field correlation function is a real valued even function of time:
\begin{equation} \label{eq:timerev_corrfunction}
\langle b_i (0) b_j(\tau)\rangle = \langle b_i (0) b_j (-\tau)\rangle,
\end{equation}
(when one of the components $b_i$ is linearly dependent on the particle velocity, as it is in the nEDM experiment \cite{Lamoreaux2005b}, Eq. (\ref{eq:timerev_corrfunction}) is no more valid).
From Eq. (\ref{eq:timeinv_corrfunction}) and (\ref{eq:timerev_corrfunction}) immediately follows that the cross-correlation terms are equal:
\begin{equation}
\langle b_i (0)b_j(\tau)\rangle = \langle b_j(0) b_i (-\tau)\rangle = \langle b_j (0)b_i (\tau)\rangle.
\end{equation}
This allows us to drop out the cross-correlation terms in Eq. (\ref{eq:long-relaxation-rate}) and (\ref{eq:frequency-shift}). 

By integrating Eq. (\ref{eq:long-relaxation-rate}) twice by parts and using the fact that $\langle b_{i}(0)b_{i}(\tau)\rangle\rightarrow0$ (with $i=x,y,z$) for $\tau\rightarrow+\infty$, we can write:

\begin{widetext}
\begin{equation}\label{eq:intermediate-equation}
\Gamma _1=\frac{\gamma ^2 }{\omega _0 ^2}\left(-\frac{{\rm{d}}}{{\rm{d}}\tau}\langle b_x(0)b_x(\tau)+b_y(0)b_y(\tau)\rangle\mid _{\tau =0} - \int _0 ^{\infty} {\rm{d}}\tau {\cos \omega _0 \tau} \frac{d^2}{d\tau ^2}\langle b_x (0)b_x(\tau ) + b_y (0)b_y (\tau )\rangle \right).
\end{equation}
\end{widetext}

We analytically treat the first term in Eq. (\ref{eq:intermediate-equation}) (for details, see Appendix A) using the definition of ensemble average, replacing the time derivative by the right side of the diffusion equation (\ref{eq:diffusion}), using the Green-Ostrogradski theorem {(also called the divergence theorem)} \cite{Appel} and applying the initial and boundary conditions. 
The result is:
\begin{equation}
\frac{\mathrm{{d}}}{\mathrm{{d}}\tau}\langle b_{i}(0)b_{i}(\tau)\rangle|_{\tau=0}=-D\overline{(\vec{\nabla}b_{i})^{2}}.%
\end{equation}
To estimate the second term of Eq. (\ref{eq:intermediate-equation}), we applied the Lebesgue-Riemann lemma \cite{Appel} (see Appendix B) which says that for high frequency (adiabatic regime), it goes to zero faster than the first term.
Finally we can write:
\begin{equation}
\Gamma_{1}\approx D\frac{\overline{\left\vert \vec{\nabla}b_{x}\right\vert ^{2}}+\overline{\left\vert \vec{\nabla}b_{y}\right\vert ^{2}}}{B_{0}^{2}.%
}\label{eq:gamma1-Adiab-Diff}%
\end{equation}
Notice that we obtain a volume average and not an ensemble average.
One can see that, for uniform gradients, our main result (\ref{eq:gamma1-Adiab-Diff}) is equal to Eq. (\ref{eq:commonlyknowresult}).
However, Eq. (\ref{eq:gamma1-Adiab-Diff}) can also be applied to arbitrary shapes of cells and magnetic field inhomogeneities.

One can carry out a similar procedure to obtain a volume average formula for the frequency shift. 
This time, integrating  {the second term in Eq. (\ref{eq:frequency-shift}) }only once by parts,  {we} obtain:
\begin{widetext}
\begin{equation}\label{eq:dw-intermadiate-equation}
\delta\omega=\frac{\gamma^{2}}{2\omega_{0}}\left(  \langle b_{x}(0)b_{x} (\tau)+b_{y}(0)b_{y}(\tau)\rangle|_{\tau=0}-\int_{0}^{\infty }\mathrm{{d}}\tau\cos{\omega}_{o}{\tau}\frac{\mathrm{{d}} }{\mathrm{{d}}\tau}\langle b_{x}(0)b_{x}(\tau)+b_{y}(0)b_{y}(\tau )\rangle\right)
\end{equation}
\end{widetext}
Again, according to the Lebesgue-Riemann lemma {and the definition of volume average (\ref{def:volume-average})}, the second term in Eq. (\ref{eq:dw-intermadiate-equation}) goes to zero as $\omega _0$ goes to infinity, leading to the following result for the high frequency asymptote of the frequency shift:
\begin{equation}
\delta\omega\approx\frac{\gamma^{2}}{2\omega_{0}}\left(  \overline{b_{x}^{2} }+\overline{b_{y}^{2}}\right).  \label{eq:FreqShift-Adiab}
\end{equation}
In the derivation of Eq. (\ref{eq:FreqShift-Adiab}), we did not use any propagator, meaning that it is valid not only for the diffusive regime but also for quasi-ballistic movements of particles.
 The only hypothesis is that the correlation function $\langle b_{i}(0)b_{i}(\tau)\rangle$ goes to zero when $\tau$ tends to infinity.

The remarkable feature of our results (\ref{eq:gamma1-Adiab-Diff}, \ref{eq:FreqShift-Adiab}) is that they were obtained without any specific assumptions on the cell geometry and on the shape of magnetic field. 
This universality of adiabatic diffusive motions explains why previous results for the longitudinal relaxation $\Gamma_{1}$ due to constant gradient magnetic fields are exactly the same for 1D models \cite{Petukhov2010}, for a 3D rectangular cell \cite{Clayton2011}, and for spheres \cite{Cates1988,McGregor1990}.

\section{Validity domain of the result and discussion}

Let us discuss the   {domain of validity}  for the main result (\ref{eq:gamma1-Adiab-Diff}). 
First, the evolution equation (\ref{eq:diffusion}) of the propagator is only valid at times much longer  {than} the collision time between two particles $\tau_{{coll}}$. 
This means that for the corresponding spectrum, the frequency $\omega_{0}$ should satisfy $\omega _{0}\ll1/\tau_{coll}$. 
A more general theory of the correlation functions  for arbitrary field and any time scale (from quasi-ballistic to diffusive motion) may be found in \cite{Swank2012}.

It is important to notice that the Lebesgue-Riemann lemma does not tell  how fast the second term in Eq. (\ref{eq:intermediate-equation}) goes to zero with $\omega_{0},$  {nor}  {the} critical frequency above which we can apply our result (\ref{eq:gamma1-Adiab-Diff}). 
To find an answer on the last point, let's first have a look  {at} the results shown on Fig. \ref{fig:comb_plot} (see also Appendix C). 
For a magnetic field perturbation $b(r)$ weakly dependent on the position within the cell (low power of power law perturbing field, Appendix C), the exact result (\ref{eq:Gamma1k}) tends to the asymptotic behavior (\ref{eq:gamma1-Adiab-Diff}) when the  {Larmor }phase accumulated during the time required by the gas
particles to diffuse across the cell $\phi_{L}=\omega_{0}\tau_{D}$, ($\tau _{D}=\frac{L^{2}}{\pi^{2}D}$), is significant: $\phi_{L}\gg1$. 
In the case of weakly position dependent inhomogeneities, $L$ is equal to the typical cell size. 
This means that the spin diffusive motion is adiabatic everywhere within the cell. 
A very similar behavior has been observed for a short-range exponential spin-dependent force $b(r)\approx b_{0}\exp (-r/\lambda)$ \cite{Petukhov2010,Swank2012} ($\tau_{\lambda}=\frac{\lambda ^{2}}{\pi^{2}D}$) and for the magnetic field of a small hard-core dipole of radius $\rho$ placed inside the spherical cell of radius $R\gg\rho$ \cite{Petukhov} ($\tau_{\rho}=\frac{\rho^{2}}{\pi^{2}D}$). 
In those cases where the inhomogeneity is localized within a region of size $\lambda$ which is much less than the effective size of the cell, the typical spatial scale is not the cell size $L$ but the magnetic inhomogeneity size $\lambda$. 
We define the size of inhomogeneity as the size of the region, $\lambda$, where the magnetic forces acting on the spins are significant. 
Notice, that in all above examples  $\tau _{\rm{corr}} = \tau_{\rm{\lambda}}$.
Those observations allow us to formulate a general condition for our result to be valid:
\begin{equation}
\frac{1}{\tau_{\lambda}}\ll\omega_{0}\ll\frac{1}{\tau_{\rm{coll}}},%
\label{eq:intermediate-frequency-condition}%
\end{equation}

Finally, the requirement $\tau_{\rm{corr}}\ll T$ for applying the Redfield theory can be translated into requirements on the strength of the magnetic inhomogeneities. 
This step depends on the experimental conditions and the observable of interest \cite{Redfield1965,Goldman2001}.
The adiabatic regime $\omega_{0} \tau_{\rm{corr}}\gg1$, for which $1/T_{1}$ is approximately equal to $\frac{\langle b^2 \rangle }{B_0^2 \tau _{\rm{corr}}}$ \cite{Goldman2001} leads us to the condition $\sqrt{\langle b^2 \rangle}\ll \vert B_{0}\vert$. 
For a local perturbation, like the field from a small size magnetic impurity, we expect a stronger limit to the strength of the perturbation $|b| \ll \vert B_0\vert $. 
In the opposite limit, there always will be regions within the cell volume, where the total field $\vec{B}=\vec{B_{0}}+\vec{b}$ crosses zero value. 
Within such regions the spin motion is no more adiabatic and our main result is not applicable (example of such situation may be found in \cite{Schmiedeskamp2006}). 
Finally, the validity domain of our result may be formulated as the following:
\begin{equation}\label{eq:frequency-condition}
\frac{1}{\tau_{\rm{\lambda}}} \ll \omega _0 \ll \frac{1}{\tau _{\rm{coll}}}\, \rm{and}\, |b| \ll \vert B_0\vert.
\end{equation}
This combined condition guarantee the validity of the Redfield theory approach and the validity of the adiabatic limit everywhere within the cell.

\section{Conclusion}

We considered the gaseous spin relaxation  of particles due to their restricted diffusive motion in an arbitrary geometry cell and exposed to 	a magnetic field, depending only on position with an arbitrary dependence.
Applying the  (perturbation) Redfield theory and the diffusion equation, we derived the asymptotic behavior (\ref{eq:gamma1-Adiab-Diff}) of the longitudinal relaxation rate for the high frequency limit (\ref{eq:frequency-condition}). 
This asymptotic behavior is found to be in excellent agreement with all previously know results \cite{Cates1988,McGregor1990,Petukhov2010,Swank2012,Petukhov}. 
We also derived a general high frequency asymptote for the frequency shift (\ref{eq:FreqShift-Adiab}), regardless of the  {type of } motion (diffusive or ballistic) in the magnetic inhomogeneity. 
Due to the generality of our results, we expect that they will find a very broad spectrum of applications.

\appendix

\section{Treatment of the leading term in Eq. (\ref{eq:intermediate-equation})}

We have to calculate the first term in Eq. (\ref{eq:intermediate-equation}):
\begin{equation}
\label{eq:A}A = \frac{\mathrm{{d}}}{\mathrm{{d}}\tau} \langle b_{i} (0)b_{i}(\tau)\rangle\vert_{\tau= 0}.%
\end{equation}
Using the correlation function definition given by Eq. (\ref{defn:corr-functions}) and the diffusion equation (\ref{eq:diffusion}), one can write:
\begin{eqnarray*}
A& = \frac{1}{V}\displaystyle\int_{V} \displaystyle\int_{V} \mathrm{{d}}\vect{r} \mathrm{{d}}\vec{r}_{0} b_{i}(\vect{r})b_{i}(\vect{r}_{0}) \frac{\mathrm{{d}}}{\mathrm{{d}}\tau}p(\vect{r},\tau\vert\vect{r}_{0})\\
& = \frac{D}{V} \displaystyle\int_{V} \mathrm{{d}}\vect{r}_{0} b_{i}(\vect{r}_{0})\int_{V} \mathrm{{d}} \vect{r} b_{i}(\vect{r}) \Delta p(\vec{r},\tau\vert\vec{r}_{0}).
\end{eqnarray*}

The Green-Ostrogradski theorem gives the following relations:%
\begin{eqnarray}
\label{eq:GO1}\int_{V} \left(  \vec{F}\cdot\vect{\nabla}g + g(\vec{\nabla}\cdot\vec{F})\right)  \mathrm{{d}}\vec{r}  & = \displaystyle\oint _{S} g\vect{F}\cdot\mathrm{{d}}\vect{S},\\
\label{eq:GO2} \int_{V} \left(  f\vec{\nabla}^{2} g + \vec{\nabla}f\cdot\vec{\nabla}g\right) \mathrm{{d}} \vec{r}  & = \displaystyle\oint _{S} f\vec{\nabla}g\cdot\mathrm{{d}}\vec{S}.%
\end{eqnarray}
According to Eq. (\ref{eq:GO1}), we have:%
\begin{eqnarray*}
B (\tau ) & =& \displaystyle\int_{V} \mathrm{{d}} \vec{r} b_{i}(\vec{r}) \vec{\nabla}\cdot\left( \vec{\nabla}p(\vec{r},\tau\vert\vec{r}_{0})\right) \\
& =& \displaystyle\oint _{S} b_{i}(\vec{r}) \vec{\nabla}p(\vec{r},\tau\vert\vec{r} _{0})\cdot\mathrm{{d}}\vec{S} - \displaystyle\int_{V} \mathrm{{d}}\vec{r}\vec{\nabla}b_{i}(\vec{r})\cdot\vec{\nabla}p(\vec{r},\tau\vert\vec{r}_{0}).
\end{eqnarray*}
Using boundary conditions given by Eq. (\ref{eq:boundaries}) and then using Eq. (\ref{eq:GO2}), we obtain:%
\begin{eqnarray*}
B (\tau ) & =& -\displaystyle\int_{V} \mathrm{{d}} \vec{r} \vec{\nabla}b_{i}(\vec{r})\cdot \vec{\nabla}p(\vec{r},\tau\vert\vec{r}_{0})\\
& =& - \displaystyle\oint _{S} p(\vec{r},\tau\vert\vec{r}_{0}) \vec{\nabla}b_{i}(\vec {r})\cdot\mathrm{{d}}\vec{S}\\
& +& \displaystyle\int_{V} \mathrm{{d}}  \vec{r} p(\vec{r},\tau\vert\vec{r}_{0})\Delta b_{i}(\vec{r}).
\end{eqnarray*}

For $\tau= 0$, the initial condition (\ref{eq:initial-condition}) gives:%
\begin{eqnarray*}
B (\tau =0) & =& -\displaystyle\oint _{S} \delta(\vec{r}-\vec{r}_{0})\vec{\nabla}b_{i}(\vec {r})\cdot\mathrm{{d}}\vec{S}\\
& +& \displaystyle\int_{V} \mathrm{{d}}\vec{r} \delta(\vec{r}-\vec{r}_{0})\Delta b_{i}(\vec{r})\\
& =& -\displaystyle\oint _{S} \delta(\vec{r}-\vec{r}_{0})\vec{\nabla}b_{i}(\vec{r})\cdot\mathrm{{d}}\vec{S} + \Delta b(\vec{r}_{0}),
\end{eqnarray*}
leading to:%
\begin{eqnarray*}
A  & = \frac{D}{V}\displaystyle\int_{V} \mathrm{{d}} \vec{r}_{0} b_{i}(\vec{r}_{0})\left(  -\displaystyle\oint _{S} \delta(\vec{r}-\vec{r}_{0})\vec{\nabla}b_{i}(\vec {r})\cdot\mathrm{{d}}\vec{S}\right) \\
& + \frac{D}{V}\displaystyle\int_{V} \mathrm{{d}} \vec{r}_{0} b_{i}(\vec{r}_{0}) \Delta b_{i}(\vec{r}_{0})
\end{eqnarray*}
\begin{equation}
\label{eq:one}= -\frac{D}{V} \displaystyle\oint _{S} b_{i}(\vec{r})\vec{\nabla}b_{i}(\vec{r})\cdot\mathrm{{d}}\vec{S} + \frac{D}{V}\int_{V} \mathrm{{d}} \vec{r}_{0} b_{i}(\vec{r}_{0}) \Delta b_{i}(\vec{r}_{0}).
\end{equation}

Another use of Eq. (\ref{eq:GO2}) on the second term in right member of Eq. (\ref{eq:one}) leads to:%
\begin{eqnarray*}
A  & = -\frac{D}{V} \displaystyle\oint _{S} b_{i}(\vec{r})\vec{\nabla}b_{i}(\vec {r})\cdot\mathrm{{d}}\vec{S} + \frac{D}{V}\displaystyle\oint _{S} b_{i}(\vec{r})\vec{\nabla}b_{i}(\vec{r})\cdot\mathrm{{d}}\vec{S}\\
& -\frac{D}{V}\displaystyle\int_{V} \mathrm{{d}}\vec{r}_{0} \vec{\nabla}b_{i}(\vec{r}_{0})\cdot\vec{\nabla}b_{i}(\vec{r}_{0}).
\end{eqnarray*}
Finally, we can write:
\begin{equation}
\frac{\mathrm{{d}}}{\mathrm{{d}}\tau}\langle b_{i}(0)b_{i}(\tau)\rangle|_{\tau=0}=-D\overline{\left\vert \vec{\nabla}b_{i}\right\vert ^{2}}\label{eq:premier-terme-resultat}.%
\end{equation}

\section{Treatment of the second term in Eq. (\ref{eq:intermediate-equation})}

Consider the second term in Eq. (\ref{eq:intermediate-equation}). 
This term corresponds to a Fourier transform of the second order time derivative of the magnetic field inhomogeneities divided by $\omega _{0}^{2}$. 
To estimate the high-frequency behavior of this term, we apply the Lebesgue-Riemann lemma which states that the cosine transform of an integrable function goes to zero as $\omega_{0}$ goes to $\infty$. 
Therefore, we have to estimate the following integral:
\begin{equation}
I=\int_{0}^{\infty}\mathrm{{d}}t|\frac{\mathrm{{d}}^{2}}{\mathrm{{d}}t^{2} }\langle b_{i}(0)b_{i}(t)\rangle|.
\end{equation}
Now, performing the integration formally, we can write:
\begin{equation}
I=\frac{\mathrm{{d}}}{\mathrm{{d}}t}\langle b_{i}(0)b_{i}(t)\rangle |_{t=0,\infty}.
\end{equation}
$\langle b_{i}(0)b_{i}(t)\rangle$ is an auto-correlation function which describes a diffusive motion and, hence, it goes to zero together with its time derivative as $t\rightarrow\infty$. 
Finally, we can write:
\begin{equation}
I=-\frac{\mathrm{{d}}}{\mathrm{{d}}t}\langle b_{i}(0)b_{i}(t)\rangle |_{t=0}\label{eq:second-term}.
\end{equation}

We see that Eq. (\ref{eq:second-term}) is the same expression which has been calculated in Eq. (\ref{eq:premier-terme-resultat}). 
Therefore, our function $\frac{\mathrm{{d}}^{2}}{\mathrm{{d}}t^{2}}\langle b_{i}(0)b_{i}(t)\rangle$ is integrable and according to Lebesgue-Riemann lemma, its cosine transform goes to zero when $\omega_{0}$ goes to $\infty$. 
Hence, the second term in Eq. (\ref{eq:intermediate-equation}) decays faster than the first one as $\omega_{0}$ goes to $\infty$.

\section{Exact solution of spin-relaxation problem due to 1D diffusive motion in inhomogeneous field of power law shape}

In this section, we will consider spin relaxation of spin 1/2 particles due to a restricted $z\in\left[  -L/2,L/2\right]  $ one-dimensional (1D) diffusive motion in a magnetic field $\vec{B}(z)=B_{0}\vec{e}_{z}+\vec{b}(z)$ where $B_{0}$ is an homogeneous part and $\vec{b}(z)=b_{x}(z)\vec{e}_{x} +b_{y}(z)\vec{e}_{y}+b_{z}(z)\vec{e}_{z}$ with $\left\Vert \vec{b}(z)\right\Vert \ll \vert B_{0}\vert$. 
Instead of considering 3D geometries, we consider a 1D problem with only one component $b(z)$. 
Then to obtain the final result, we will need to make a proper combination of the components. 
To be more specific, we will consider the perturbation field $b(z)$ which has the form of a power law:
\begin{equation}
b(z) = b_{0} \left(  z^{k} - \frac{1}{L} \int_{-L/2} ^{L/2} z^{k} \mathrm{{d}}z\right),
\end{equation}
with $k=1,2,3,4$. 
Notice that with this definition, the average of the volume perturbation is null. 
This type of perturbation is of great practical importance since many ideal magnetic systems designed to create homogeneous fields have the dominant component of inhomogeneity of the shape with $k = 2$.
Moreover, for a popular Helmholtz pair of DC coils or End-compensated solenoid one can expect $k=4$. 
Imperfections in the wiring will give terms with $k=1,3...$. 
The solution of the problem for $k=1$ is well known and will be given here for completeness.

A general expression for longitudinal relaxation $\Gamma_{1}$ due to 1D diffusive motion reads \cite{Petukhov2010,Clayton2011,Swank2012}:
\begin{widetext}
\begin{equation}
\label{eq:gamma1_1D}
\Gamma _1 = 2\gamma ^2 \left(  \sum _{n=0} ^{\infty} \frac{\tau _{2n+1}}{1+(\omega_0 \tau_{2n+1})^2}\vert b_{2n+1} \vert ^2 + \sum _{n=1} ^{\infty} \frac{\tau _{2n}}{1+(\omega_0 \tau_{2n})^2}\vert b_{2n} \vert ^2   \right),
\end{equation}
\end{widetext}
with $\tau_{n} =\frac{\tau_{0}}{n^{2}}$ is a time constant for $n^{\mathrm{{th}}}$ order diffusive mode, $\tau_{0} = \frac{L^{2}}{\pi^{2} D}$ the time constant for the lowest mode and $D$ the diffusion coefficient. 
The $b_{2 n+1}$ and $b_{2 n}$ represents the odd and even Fourier components of the field $b(z)$:%

\begin{eqnarray}
b_{2n}  & =\frac{1}{L}\int_{-L/2}^{L/2}b(z)\cos\frac{2n\pi z}{L}\mathrm{{d}}z,\\
b_{2n+1}  & =\frac{1}{L}\int_{-L/2}^{L/2}b(z)\sin\frac{(2n+1)\pi z}{L}\mathrm{{d}}z.
\end{eqnarray}
Depending on the symmetry (odd, even) of $b(z)$, only one of the sums in Eq. (\ref{eq:gamma1_1D}) contributes to the relaxation.

With  {these} remarks, the calculation of the relaxation rate is straight-forward:
\begin{equation}\label{eq:Gamma1-k}
\Gamma_{1}^{k}=2\gamma^{2}\tau_{0}\langle g_{k}^{2}\rangle L^{2}S_{k},
\end{equation}
where $\langle g_{k}^{2}\rangle=\frac{1}{L}\int_{-L/2}^{L/2}\left( \frac{\mathrm{{d}}b(z)}{\mathrm{{d}}z}\right)  ^{2}\mathrm{{d}}z$, 
\begin{widetext}
\begin{eqnarray}
\label{eq:S1} S_1 &=& \frac{1}{2\pi ^2 \phi_L ^2}\left( 1-\frac{1}{\pi \sqrt{\phi _L /2}}\frac{\sinh \pi\sqrt{\phi _L/2}+\sin \pi\sqrt{\phi _L/2}}{\cosh \pi\sqrt{\phi _L/2} + \cos \pi\sqrt{\phi _L/2}}\right),\\
\label{eq:S2} S_2 &=& \frac{1}{2\pi ^2 \phi_L ^2}\left( 1-\frac{3}{\pi \sqrt{\phi _L /2}}\frac{\sinh \pi\sqrt{\phi _L/2}-\sin \pi\sqrt{\phi _L/2}}{\cosh \pi\sqrt{\phi _L/2} - \cos \pi\sqrt{\phi _L/2}}\right),\\
\label{eq:S3}S_3 &=& \frac{1}{2\pi ^2 \phi _L ^2}\left( 1- \frac{320}{\pi ^4 \phi _L^2}-\frac{5\sqrt{2}}{\pi ^5 \phi _L ^{5/2}}\frac{ (\pi ^4\phi _L ^2-16\pi ^2\phi _L-64)\sinh \pi\sqrt{\phi _L/2} + (\pi ^4\phi _L^2 +16\pi ^2\phi _L-64)\sin \pi\sqrt{\phi _L/2}}{\cosh \pi\sqrt{\phi _L/2} + \cos \pi\sqrt{\phi _L/2}}\right),\\
\label{eq:S4}S_4 &=& \frac{1}{2\pi ^2 \phi _L ^2}\left( 1- \frac{2688}{\pi ^4 \phi _L^2}-\frac{7\sqrt{2}}{\pi ^5 \phi _L ^{5/2}}\frac{ (\pi ^4\phi _L ^2-48\pi ^2\phi _L-576)\sinh \pi\sqrt{\phi _L/2} - (\pi ^4\phi _L^2 +48\pi ^2\phi _L-576)\sin \pi\sqrt{\phi _L/2}}{\cosh \pi\sqrt{\phi _L/2} + \cos \pi\sqrt{\phi _L/2}}\right).
\end{eqnarray}
\end{widetext}
where $\phi_{L} = \gamma B_{0} \frac{L^{2}}{\pi^{2} D}$ is the Larmor phase accumulated by the particle spin during the interval of time required to diffuse over the distance $L$. 
In the adiabatic regime, when $\phi_{L} \gg1$, all hyperbolic functions dominate over the trigonometric ones which allows us to write:%
\begin{equation}
\label{eq:Gamma_1_k-result}S_{k} \approx\frac{1}{2\pi^{2} \phi_{L}^{2}}\left( 1+ O\left(  1/\sqrt{\phi_{L}}\right) \right),
\end{equation}
and for $k=1,2,3,4$, the longitudinal relaxation rate may be written as:
\begin{equation}
\label{eq:Gamma1k}\Gamma_{1} ^{k} = D\frac{\langle g_{k} ^{2}\rangle}{B_{0}^{2}}\left(  1 + O\left(  1/\sqrt{\phi_{L}}\right) \right).
\end{equation}
The normalized frequency dependence 
\begin{equation}
\tilde{\Gamma}_{1}^{k}=\frac{\Gamma_{1}^{k}}{\gamma^{2}\tau_{0}\langle
g_{k}^{2}\rangle L^{2}}\label{eq:gamma_tild}%
\end{equation}
of the longitudinal relaxation rate for different kinds of inhomogeneities ($k=1,2,3,4$) are shown on Fig. \ref{fig:comb_plot}.

Since the transverse relaxation rate (\ref{eq:trans-relaxation-rate}) is given by:
\begin{equation} \label{eq:Gamma2_phiL}
\Gamma_{2}(\phi_{L})=\frac{1}{2}\Gamma_{1}(\phi_{L})+\gamma ^2  S_{zz}(0),
\end{equation}
and $S_{ii}(\phi _L) \ll S_{ii} (0)$, we obtain $\Gamma _2 \approx \gamma ^2 S_{zz} (0)$ in the adiabatic limit.
Eq. (\ref{eq:Gamma1-k}), (\ref{eq:Gamma2_phiL}) and Fig. \ref{fig:comb_plot} imply that the spectrum of the field correlation functions, and, hence, $T_2$  are independent of magnetic field strength $B_0$ but strongly depend on the shape of inhomogeneities, the cell size and geometry.

\bibliography{article}

\end{document}